\documentclass{aa}
\usepackage[varg]{txfonts}
\usepackage{graphicx}
\usepackage{natbib}
\usepackage{multirow}
\usepackage[colorlinks,linkcolor=blue,citecolor=blue,urlcolor=blue]{hyperref}

\begin{document}

\title{A mapping method of age estimation for binary stars: Application to the $\alpha$~Centauri system A and B}

\author{F. Thévenin\inst{1} \and
        V.~A. Baturin\inst{2} \and
        A.~V. Oreshina\inst{2} \and
        P. Morel\inst{1} \and
        S.~V. Ayukov\inst{2} \and
        L. Bigot\inst{1} \and
        A.~B. Gorshkov\inst{2}}

\institute{Université Côte d'Azur, Observatoire de la Côte d'Azur, CNRS, Laboratoire Lagrange, France \\
\email{frederic.thevenin@oca.eu}
\and
Lomonosov Moscow State University, Sternberg Astronomical Institute, Moscow, Russia}

\date{Received 3 December 2025 / Accepted 20 January 2026}

\abstract{}{}{}{}{}
\abstract
{Binary and multiple star systems are abundant, and they are important for determining the ages of exoplanet systems and Galactic populations. Given the wealth of data provided by Gaia and the upcoming PLATO mission, it is essential to improve stellar models to obtain accurate stellar ages.  Multiple systems reduce the degeneracy in the stellar parameters that control evolution, and they allow better constraints on the physical processes at work in stellar interiors.}
{Our objective is to apply a mapping technique to estimate the age of a system and the initial chemical composition. We also evaluate the influence of observational uncertainties in mass and heavy-element mixtures on results.}
{We applied an inverse calibration method to the evolution of a multiple stellar system, assuming that the stars share the same age and initial chemical composition.
This approach determines age, the initial mass fractions of helium ($Y_{\mathrm{ini}}$) and heavy elements ($Z_{\mathrm{ini}}$), as well as the convective mixing-length parameters ($\alpha_{\mathrm{A}}$ and $\alpha_{\mathrm{B}}$). It uses the observed luminosities ($L_A$ and $L_B$), radii ($R_A$ and $R_B$), and surface chemical compositions ($Z/X_A$ and $Z/X_B$).}
{We used the most recent observational data for $M$, $R$, $L$, and [Fe/H] of $\alpha$ Centauri A and B as input data for our method. We compared two assumptions for the $Z/X$ ratio, following the results for the solar composition. For an assumed high solar $Z/X_{\odot} = 0.0245$, we obtain an age of $7.8\pm0.6$ Ga, $Y_{\mathrm{ini}} = 0.284\pm0.004$, and $Z_{\mathrm{ini}}=0.0335\pm0.0015$. For a low solar $Z/X_{\odot} = 0.0181$, the derived age is $8.7\pm0.6$ Ga, $Y_{\mathrm{ini}} = 0.267\pm0.008$, and $Z_{\mathrm{ini}} = 0.025\pm0.002$. Observational errors in the stellar masses  of $\pm0.002$ lead to an age error of 0.6 Ga. Overshooting of $0.05-0.20H_p$ at the boundary of the convective core increases the age by $0.6-2.1$ Ga.
Models with higher $Z/X$ and radiative cores, with ages of $7.2-7.8$ Ga, appear preferable and show better agreement with the observed asteroseismic frequencies.}
{}


\maketitle

\section{Introduction}\label{sec:introduction}

Studies of stars in binary systems have been carried out for decades using a variety of methods and observational data with differing levels of precision and constraint. Compared to single stars, binaries offer notable advantages, most importantly the ability to impose tighter constraints on stellar ages and to test the uniqueness of calibrated stellar models.

A well-known example is the multiple system $\alpha$ Centauri \citep{morel2000}, one of the first binaries to be calibrated with a modern approach. The $\alpha$ Centauri system consists of $\alpha$ Centauri A and B, two Sun-like stars in a close binary orbit, and $\alpha$ Centauri C (Proxima Centauri), a low-mass red dwarf. Their proximity and solar-like nature make $\alpha$ Centauri A and B benchmark stars for calibrating stellar evolution models. The binary configuration enables robust determination of age and composition, making the system ideal for testing age-dating tools used in future studies of Galactic populations.

Another binary star, $\zeta$ Herculis, was calibrated before $\alpha$~Centauri \citep{chmielewski1995}.
The authors applied evolutionary tracks simultaneously to both components and predicted asteroseismic frequencies for comparison with observations. They concluded that the system remained insufficiently constrained, particularly in stellar masses and parallax. A subsequent recalibration aimed to improve upon this earlier analysis \citep{morel2001}.

With the advent of high-precision parallaxes from the Gaia mission, particularly for numerous binary systems with accurately measured masses and for multiple systems in clusters, it is now timely to revisit such targets to test and refine calibration methods. Expanding the analysis to both binaries and higher-order multiples will enable more precise stellar age determinations, thereby improving the accuracy of age tagging for stellar populations across the Milky Way.

This work considers the development of a numerical method for calibrating multiple systems. We used recent high-precision measurements of stellar masses, luminosities, and radii, which are essential inputs for binary modeling. Our goal was to create an automated tool to determine the age and initial chemical composition,  specifically the initial helium mass fraction ($Y_{\mathrm{ini}}$) and heavy-element mass fraction ($Z_{\mathrm{ini}}$), of binary star systems. The method uses observational constraints such as stellar masses, radii, luminosities, and surface metallicity ratios (Z/X). A central assumption is that both stars formed simultaneously and therefore share the same age and initial composition.

We aimed to address key questions: Does a physically consistent solution exist for the given constraints? If so, is the solution unique? How do observational uncertainties and the assumed heavy-element mixture influence the results?

For broad applicability, we did not include seismic constraints in our general calibration, as asteroseismic frequency data for binaries remain scarce, despite their demonstrated ability to significantly improve age estimates (\cite{joyce2018}). However, for the test on $\alpha$ Centauri, we used available seismic data to assess our solution, while noting that the number of detected frequencies is still limited.

In Sect.~\ref{sec:method}, we develop the inversion mapping method. We test our tool on the $\alpha$ Centauri system in Sect.~\ref{sec:Application}. Our conclusions are summarized in Sect.~\ref{sec:conclusion}.

\begin{table*}
\centering
\caption{Results from recent observations of $\alpha$ Centauri A and B.}
\label{tab:observations}
\begin{tabular}{lcccc}
\hline\hline
 & \multicolumn{2}{c}{$\alpha$ Centauri A} & \multicolumn{2}{c}{$\alpha$ Centauri B} \\
 \hline
 & \citet{kervella2017} & \citet{akeson2021} & \citet{kervella2017} & \citet{akeson2021} \\
$M/M_{\odot}$ & $1.1055\pm0.0039$ & $1.0788\pm0.0029$ & $0.9373\pm0.0033$ & $0.9092\pm0.0025$ \\
$R/R_{\odot}$ & $1.2234\pm0.0053$ & $1.2175\pm0.0055$ & $0.8632\pm0.0037$ & $0.8591\pm0.0036$ \\
$L/L_{\odot}$ & $1.521\pm0.015$ & $1.5059\pm0.0019$ & $0.503\pm0.006$ & $0.4981\pm0.0007$ \\
\hline
 & \citet{porto2008} & \citet{morel2018} & \citet{porto2008} & \citet{morel2018}\\
$[Fe/H]^{*}_{\odot}$ & $0.24\pm0.03$ & $0.237\pm0.007$ & $0.25\pm0.04$ & $0.221\pm0.016$ \\
\hline
$Z/X$ & $0.039\pm0.006^a$ & $0.0423\pm0.0050^b$ & $0.039\pm0.006^a$ & $0.0407\pm0.0050^b$ \\
 &  & $0.0312\pm0.0050^c$ &  & $0.0300\pm0.0050^c$ \\
\hline
\end{tabular}
\tablefoot{
\tablefoottext{a}{$Z/X$ is computed using $(Z/X)_{\odot}=0.022$ (see \citealt{joyce2018})},
\tablefoottext{b}{$Z/X$ is computed using $(Z/X)_{\odot}=0.0245$, as in the high-Z Sun \citep{grevesse1993}},
\tablefoottext{c}{$Z/X$ is computed using $(Z/X)_{\odot}=0.0181$, as in the low-Z Sun \citep{asplund2009}}.
}
\end{table*}

\section{The inverse mapping method for stellar calibration }
\label{sec:method}

\subsection{Mathematical formulation of the problem}\label{subsec:MathFormulation}

We present a mathematical formulation of the problem of modeling stars A and B in a binary system. The key concept is a stellar evolutionary track as a mapping from one space to another. The evolutionary track is defined as a sequence of models for a star with mass $M_j$, which maps the three initial parameters of the star $\{\alpha_j,Y_{\mathrm{ini}}, $ and $  Z_{\mathrm{ini}}\}$ to the three observed parameters $\{R,L$, and $Z/X\}_j$ at each assumed stellar age $t_s$ Here, the index $j$ refers to star A or B. We define $Y_{\mathrm{ini}}$ as the initial helium mass fraction, $Z_{\mathrm{ini}}$ as the initial heavy-element mass fraction (i.e., elements heavier than helium), $\alpha_j$ as the mixing-length parameter, $R$ as the stellar radius, $L$ as the luminosity, and $Z/X$ as the ratio of heavy-element to hydrogen mass fractions. 

The idea of evolutionary mapping originates from \cite{noels1991}. An extensive review of its applications is provided by \cite{guenther2000}.  
Mathematically, direct evolutionary mapping is expressed as
\begin{equation}
E(M_j): \left[ \{\alpha_j,Y_{\mathrm{ini}},Z_{\mathrm{ini}}\} \xrightarrow{t_s} \{R,L,Z/X\}_j \right],
\label{eq:dir_map}
\end{equation}

\noindent where $t_s$ is the stellar age and serves as a parameter of the mapping.  This formulation represents a set of evolutionary tracks as functions of age, specific to each stellar component. We analyze the mapping by examining these evolutionary tracks across the relevant parameter space.

Our goal is to identify, among all possible tracks, a pair of evolutionary tracks 
$E_A(t)$ and $E_B(t)$ that correspond to the given observed parameters $\{R,L,Z/X\}_{A/B}^{\mathrm{observed}}$ 
of stars A and B. The resulting tracks must satisfy the binary constraints: they must have the same age ($t_A = t_B$), and also have the same initial chemical composition 
$\{Y_{\mathrm{ini}},Z_{\mathrm{ini}}\}_A = \{Y_{\mathrm{ini}},Z_{\mathrm{ini}}\}_B. $
To solve the problem, we introduce an inverse mapping $E^{-1}$:
\begin{equation}
E^{-1}(M_j): \left[ \{R,L,Z/X\}_j^{\mathrm{observed}} \xrightarrow{t_s} \{\alpha_j,Y_{\mathrm{ini}},Z_{\mathrm{ini}}\} \right].
\label{eq:inv_mapping}
\end{equation}
In Eq.~(\ref{eq:inv_mapping}), we do not require equality of the convection parameters $\alpha_A$ and $\alpha_B$. The values of $\alpha_A$ and $\alpha_B$ can differ. We do not impose prior assumptions on these parameters.

When considering a single star, we have three known parameters ($R$, $L$, and $Z/X$) and three unknown ($\alpha_j$, $Y_{\mathrm{ini}}$, and $Z_{\mathrm{ini}}$); the problem is determined for every fixed time $t_s$. For a pair of stars, there are six known parameters ($R_A$, $L_A$, $(Z/X)_A$, $R_B$, $L_B$, and $(Z/X)_B$) and five unknowns ($t_s$, $Y_{\mathrm{ini}}$, $Z_{\mathrm{ini}}$, $\alpha_A$, and $\alpha_B$). The problem is overdetermined. The observed parameters $\{R,L,Z/X\}_{A/B}^{\mathrm{observed}}$ naturally include measurement uncertainties. 
Assessing the impact of these observational errors is fundamental; we address this in our analysis of $\alpha$ Cen. However, when defining the stellar evolutionary tracks and formulating the binary system constraints, we do not include these errors.

\subsection{Inverse mapping method}\label{subsec:Method}

Computing an inverse mapping in explicit form is challenging. We used sequential iterations and inverted the differential of the evolutionary map Eq.~(\ref{eq:dir_map}).

For a single star, the mapping provides a one-to-one correspondence between two 3D parameter spaces. It is differentiable with respect to the three input parameters $P_{\mathrm{ini}} = \{\alpha_j,Y_{\mathrm{ini}},Z_{\mathrm{ini}}\}$. Therefore, we constructed the differential of the mapping via the Jacobian matrix, which captures the sensitivity of the observable parameters to variations in the initial stellar parameters,
\begin{equation}
J_E(t_s) = \frac{\partial \{R,L,Z/X\}}{\partial \{\alpha,Y_{\mathrm{ini}},Z_{\mathrm{ini}}\}}.
\label{eq:Jacobian}
\end{equation}

The Jacobian $J_E$ is a matrix 3 × 3 that depends on the assumed age of the system. It must be non-singular, i.e., its determinant must be nonzero, $\mathrm{det}(J_E)\neq 0$. If this condition is not fulfilled, the inverse mapping has no solutions or infinitely many solutions.

The properties of the Jacobian are straightforward. It allows us to determine how much the observed (target) parameters $T = \{R,L,Z/X\}$ of the star change in response to a variation in the initial parameters by a vector $\delta P_{\mathrm{ini}}$, such that
\begin{equation}
\Delta T = J_E \cdot \delta P_{\mathrm{ini}}.
\label{eq:dT}
\end{equation}

\noindent The dot in Eq.~(\ref{eq:dT}) denotes matrix multiplication.

The core principle of our method is that the search for the inverse mapping $E^{-1}$ is replaced by the calculation of the inverse Jacobian $J_E^{-1}$. This inverse matrix enables an efficient determination of the initial stellar parameters that produce the observed parameters $T$. The process is iterative. If the difference between the observed parameters of the currently constructed evolutionary track, $E^{(i)}(t_s)$, and the target point, $T$, is $\Delta_i T = T - E^{(i)}(t_s)$,  we use as the next approximation for the track $i + 1$:
\begin{equation}
P_{\mathrm{ini}}^{i+1} = P_{\mathrm{ini}}^i + J_E^{-1} \cdot \Delta_i T.
\label{eq:Pini}
\end{equation}

The convergence of the iterative process is generally rapid: typically, around 12 iterations are sufficient for star A, and only about five iterations are needed for star B.

The inverse Jacobian method facilitates the study of the influence of observational errors on the evolutionary track. It is sufficient to estimate the perturbations of the observational quantities $\Delta T^{\mathrm{obs}}$ and use Eq.~(\ref{eq:Pini}), to determine the effect of observation errors on the track parameters. The inverse Jacobian $J_E^{-1}$ depends weakly on the desired track parameters. We computed $J_E$ for a sequence of ages, but used only the closest in age for the calculations.

Calculations of suitable parameters $P_{\mathrm{ini}}^{(0)}(t_s)$ were carried out for the entire series of ages $t_s$. As a result, we obtained a sequence of initial parameters $\{\alpha(t_s),Y_{\mathrm{ini}}(t_s),Z_{\mathrm{ini}}(t_s)\}$ under which the mapping $E$ should fall within the given values of the observed parameters $T$. If necessary, the approximations for $P_{\mathrm{ini}}^{(j)}(t_s)$ can be refined by applying Eq.~(\ref{eq:Pini}) and calculating the corresponding tracks.

The inverse calibration method is most suitable for solar-type stars on the main sequence. For older stars, applying the method appears problematic. For late spectral type stars with masses below 0.8 $M_\odot$, further study is required. Estimating critical uncertainty levels for mass, radius, and luminosity, which render the method ineffective, is generally difficult and requires case-by-case verification. Possible candidates for applying the method include $\alpha$ Centauri, 16 Cygni, and similar stars.

\section{Application to $\alpha$ Centauri A and B}
\label{sec:Application}

\subsection{Input data}\label{subsec:InputData}

As mentioned previously, the nearby $\alpha$ Centauri system is a well-studied triple system, comprising the close binary $\alpha$~Centauri A and B, which are two Sun-like stars, and  $\alpha$ Centauri C (Proxima Centauri), a low-mass red dwarf. Their proximity to the Sun and solar-like properties make $\alpha$ Centauri A and B benchmark stars for calibrating stellar evolution models. Their binary configuration enables more robust determination of age and initial composition (we account for the diffusion of elements over time in the computation of evolutionary tracks), making them ideal for testing tools used to estimate stellar ages in different Galactic populations.
Table~\ref{tab:observations} provides a summary of representative observational data. For example, \citet{kervella2017} derived precise values using the VLTI/PIONIER\footnote{ The Very Large Telescope Interferometer / Precision Integrated-Optics Near-infrared Imaging ExpeRiment.}  optical interferometer, which are used in modeling studies by \citet{joyce2018} and \citet{manchon2024}. In contrast, \citet{akeson2021}, using the Atacama Large Millimeter/submillimeter Array (ALMA) as part of a planet-detection program via differential astrometry, report notably lower values for stellar mass, radius, and luminosity than those of \citet{kervella2017}. These discrepancies highlight the importance of cross-validating observational datasets when constructing accurate models of stellar evolution. We tested our model using these binary stars. 

For our analysis, we used data from \citet{akeson2021} and \citet{morel2018} (Table~\ref{tab:observations}).
We assumed that the stellar masses were known. We adopted $M_A = 1.0788 M_{\odot}$ and $M_B = 0.9092 M_{\odot}$. 
We analyze the impact of mass uncertainties in Sect.~\ref{subsubsec:uncertainties_LRM} and \ref{subsubsec:choisemass}. The observations provide the metallicity [Fe/H], from which the ratio $Z/X$ is derived:
\begin{equation}
\left(\frac{Z}{X}\right)_* = \left(\frac{Z}{X}\right)_\odot \cdot 10^{[\mathrm{Fe}/\mathrm{H}]^{*}_{{\odot}}}.
\end{equation}

\noindent We computed $\left(Z/X\right)_*$ using a solar ratio $(Z/X)_\odot = 0.0245$ from \citet{grevesse1993}, which corresponds to a high-Z mixture. This solar $(Z/X)_\odot$ remains a subject of debate (see, for example, the review of \citealt{christensen2021}). Acceptance of a low-Z solar mixture $(Z/X)_\odot = 0.0181$ \citep{asplund2009} would alter the calibration results, although the observed [Fe/H] remains fixed. We discuss this sensitivity in Sect.~\ref{subsubsec:solarabund}.

\subsection{The CESAM2k stellar evolution code}
\label{subsec:CESAM}

We performed all computations using the 1D stellar evolution code CESAM2k \citep{morel2008, manchon2025}. We treated convection using the mixing-length theory \citep{bohm1958}. We calculated microscopic diffusion with coefficients based on Burgers equations \citep{burgers1969}, accounting for radiative acceleration. We used the OPAL 2005 equation of state \citep{rogers2002} and used the OPAL package for opacity calculations \citep{iglesias1991}. Thermonuclear reaction rates follow the NACRE compilation \citep{angulo1999}. We modeled the atmosphere using the fully radiative Eddington law \citep{eddington1930}. 
We initialized the evolutionary models at the pre-main sequence stage and assumed constant stellar masses throughout the evolution.

\subsection{Stellar modeling results}
\label{subsec:results}

\begin{figure}
\centering
\includegraphics[width=\columnwidth]{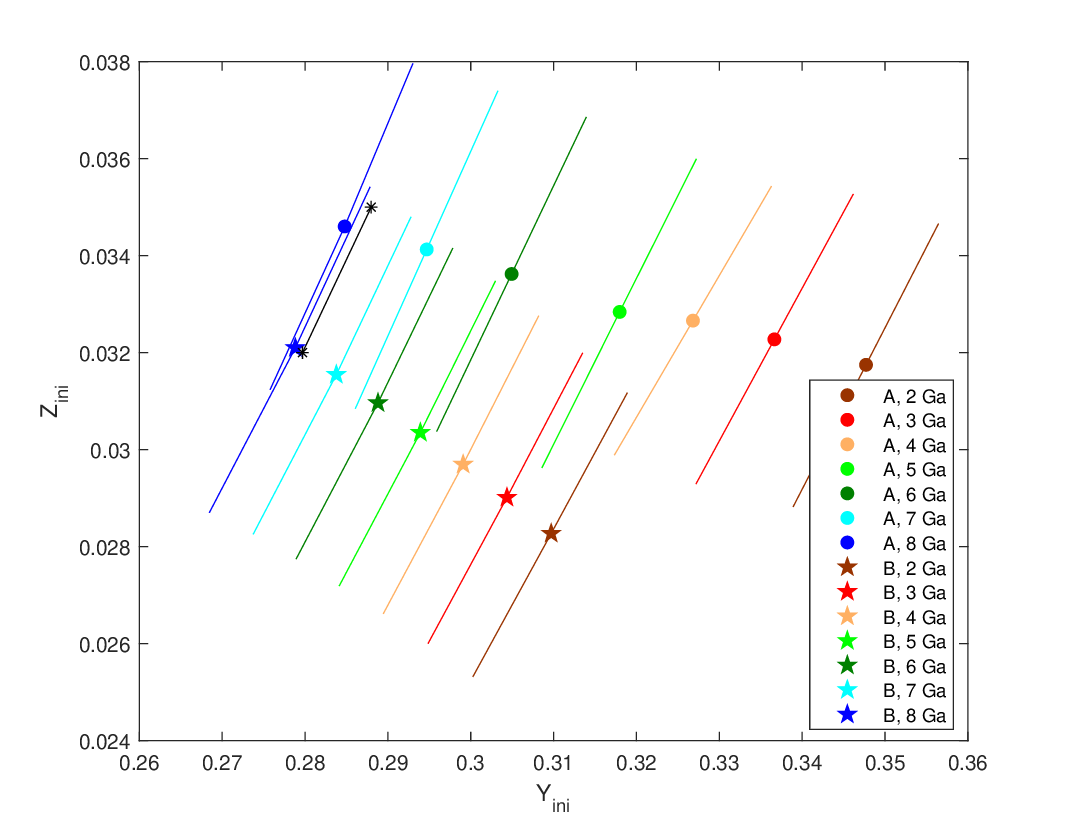}
\caption{Inverse mapping results. Initial conditions for stars A (circles) and B (stars) reproduce the observed parameters $(R,L,Z/X)_{\mathrm{obs}}$ at given ages. Colors indicate the ages, as shown in the legend. Colored lines show the initial chemical compositions obtained by accounting for observation errors $\pm\Delta(Z/X)$. The black line shows the common initial conditions $(Y_{\mathrm{ini}},Z_{\mathrm{ini}})$ for which the stars achieve the observed values $(R,L,Z/X\pm\Delta(Z/X))_{\mathrm{obs}}$ at ages = 7.801--7.826 Ga.}
\label{fig:inverse}
\end{figure}

First, we computed evolutionary tracks with masses $M_A$ and $M_B$ for sets of initial conditions $(\alpha_{\mathrm{MLT}}\pm\Delta\alpha, Y_{\mathrm{ini}}\pm\Delta Y,Z_{\mathrm{ini}}\pm\Delta Z)$ to obtain the Jacobian Eq.~(\ref{eq:Jacobian}). 
Second, we performed the inverse mapping using Eq.~(\ref{eq:Pini}). Figure~\ref{fig:inverse} 
shows the results of the inverse mapping. It presents the initial compositions of stars A and B that allow them to reach the observed parameters $(R,L,Z/X)_{\mathrm{obs}}$ at a given age, with the ages indicated in the legend. 

The data for stars A and B trace two distinct curves in the $(Y_{\mathrm{ini}}, Z_{\mathrm{ini}})$ plane. To match the observable parameters at an earlier age (e.g., 2 Ga), a higher initial helium abundance, $Y_{\mathrm{ini}}$,  and a lower initial metallicity, $Z_{\mathrm{ini}}$, are required compared to  later ages (e.g., 8 Ga). Figure~\ref{fig:inverse} shows that stars A and B share no common initial conditions $(Y_{\mathrm{ini}}, Z_{\mathrm{ini}})$ for any system age.

Fig.~\ref{fig:inverse} also illustrates how observational uncertainties affect the inferred values of $Y_{\mathrm{ini}}$ and $Z_{\mathrm{ini}}$. The impact of uncertainties in luminosity and radius is not visible at the scale of the figure.
Uncertainties of $\pm\Delta(Z/X)$ result in segments rather than single points, reflecting the range of possible initial conditions.
For young stars A and B, the segments corresponding to the same age (i.e., the same color) are well separated. For example, the two brown segments, corresponding to a stellar age of 2 Ga, are separated by $\Delta Y_{\mathrm{ini}} \approx 0.04$. The two red segments, which correspond to an age of 3 Ga, are separated by $\Delta Y_{\mathrm{ini}} \approx 0.03$. 
Separation decreases with age; for example, the cyan segments corresponding to 7 Ga are separated by $\Delta Y_{\mathrm{ini}} \approx 0.005$. At a fixed $Z_{\mathrm{ini}}$, star B has a larger $Y_{\mathrm{ini}}$ than star A. The blue segments (8 Ga) are very close to each other, with star B having a larger $Y_{\mathrm{ini}}$ than star A at fixed $Z_{\mathrm{ini}}$. Therefore, points of common initial conditions between stars A and B begin to appear at ages between 7 and 8 Ga.

We next examine the $(Y_{\mathrm{ini}},Z_{\mathrm{ini}})$ range within the $7-8$~Ga interval to identify the initial conditions under which both stars simultaneously match the observed values $(R,L,Z/X\pm\Delta(Z/X))_{\mathrm{obs}}$ at the same age. 
First, we fixed $Z_{\mathrm{ini}}$, and for a given initial $Y_{\mathrm{ini}}$, we selected $\alpha_{\mathrm{MLT}}$ such that the model falls within the observed ranges of luminosity and radius. This condition is satisfied at a specific stellar age. As illustrative cases, we chose $Z_{\mathrm{ini}} = 0.0320$, which lies near the lower limit for star A, and $Z_{\mathrm{ini}} = 0.0350$, near the upper limit for star B.

\begin{figure}
\centering
\includegraphics[width=\columnwidth]{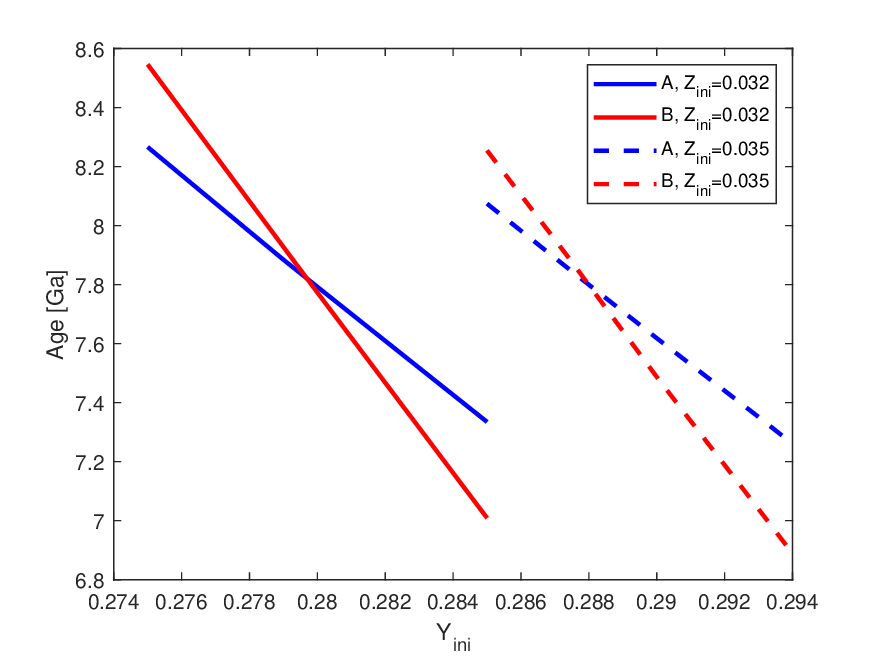}
\caption{Stellar ages as a function of $Y_{\mathrm{ini}}$ for $Z_{\mathrm{ini}}=0.032$ (solid lines) and 0.035 (dashed lines).}
\label{fig:age_yini}
\end{figure}

Figure~\ref{fig:age_yini} shows the stellar ages as a function of $Y_{\mathrm{ini}}$ for two metallicities, $Z_{\mathrm{ini}}=0.0320$ (solid lines) and 0.0350 (dashed lines). For $Z_{\mathrm{ini}} = 0.0320$ and $Y_{\mathrm{ini}} = 0.2750$, star A reaches its observed radius and luminosity at the age of 8.266 Ga, and star B at 8.605 Ga. As $Y_{\mathrm{ini}}$ increases, the stellar ages decrease. 

The slopes of the Age($Y_{\mathrm{ini}}$) curves differ between stars A and B, leading to an intersection point where both stars reach their observed properties at the same age. For $Y_{\mathrm{ini}} = 0.2796$, this common age is 7.826 Ga, representing a consistent solution for both stars. For $Z_{\mathrm{ini}} = 0.0350$, the corresponding age at the intersection point is 7.801 Ga.

\begin{figure}
\centering
\includegraphics[width=\columnwidth]{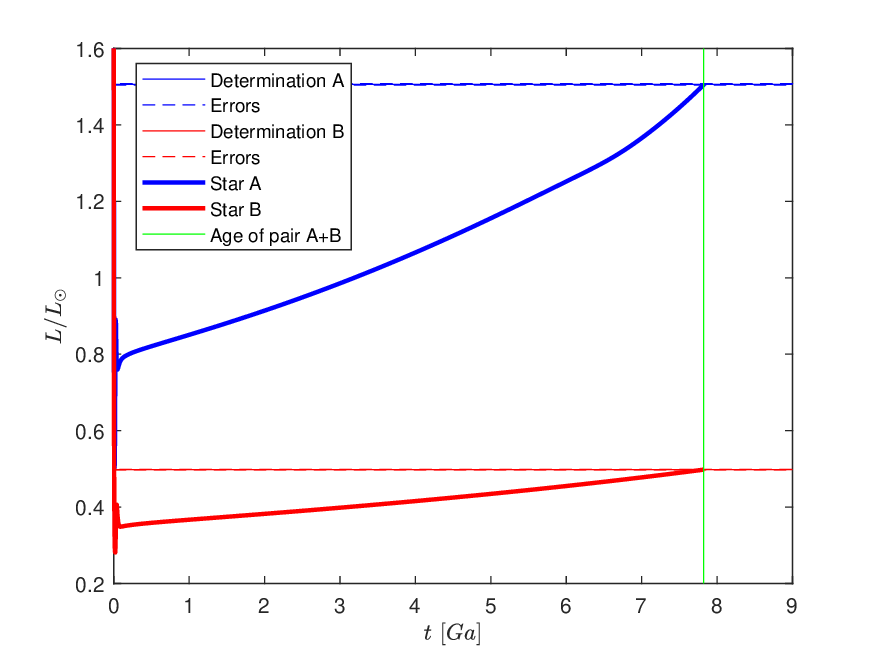}
\includegraphics[width=\columnwidth]{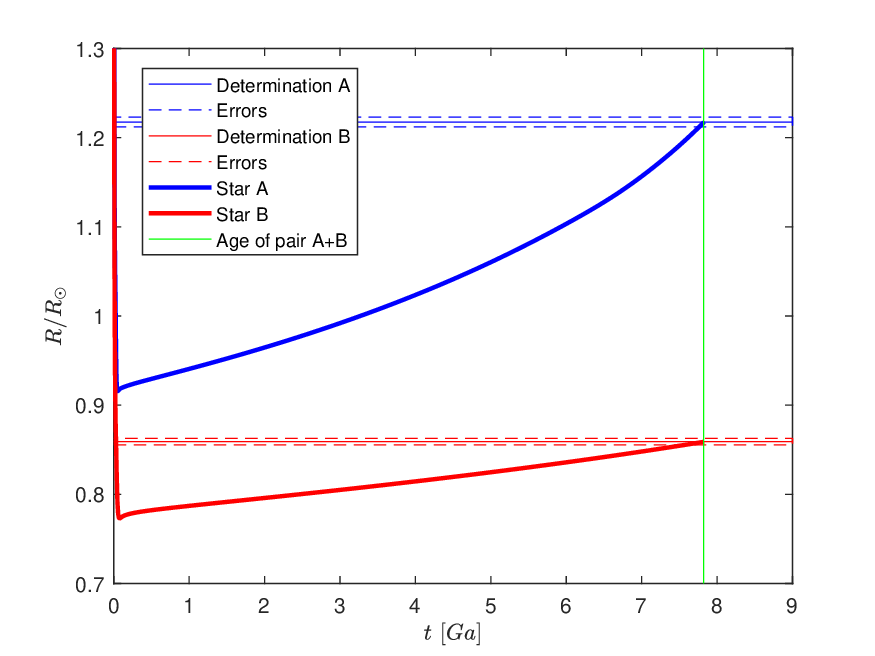}
\includegraphics[width=\columnwidth]{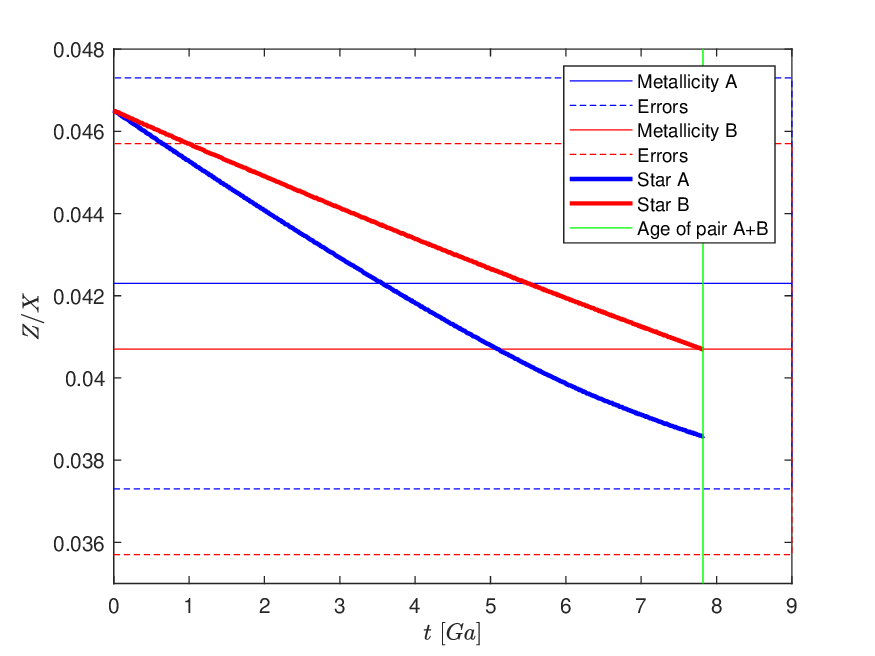}
\caption{Examples of evolutionary tracks of stars A and B for luminosity (upper panel), radius (middle panel), and $Z/X$ (lower panel), all with $Z_{\mathrm{ini}} = 0.0320$. The age of the pair is 7.826 Ga, corresponding to the intersection point in Fig.~\ref{fig:age_yini}.}
\label{fig:tracks}
\end{figure}

The two identified solutions appear as black asterisks in Fig.~\ref{fig:inverse}. Other possible pairs lie between these two limits (black segment in Fig.~\ref{fig:inverse}) and slightly beyond them. Table~\ref{tab:results_main} summarizes the corresponding initial conditions, mixing-length parameters, and stellar ages for these models. Figure~\ref{fig:tracks} presents an example of the evolutionary tracks for $Z_{\mathrm{ini}} = 0.032$. 

In the bottom panel, the evolution of $Z/X$ for star A does not reach the observed metallicity $Z/X=0.0423$ at 7.826 Ga. This occurs because no exact solution reproduces both stars within the exact $Z/X$ values given in \citep{morel2018}. However, our problem formulation seeks to find a solution that satisfies modern observations within the error limits $\pm\Delta Z/X$. Our solution satisfies this condition.

Notably, the metal-to-hydrogen mass fraction $Z/X$ decreases more rapidly during the evolution of star A than during that of star B. 
The rate of settling of heavy elements from the convective zone in solar-like stars depends on the relative mass of the convective zone, $M_{\mathrm cz}/M_*$. The larger the mass of the convective zone, the smaller the relative fraction of precipitated metals. For star A, the relative mass of the convective zone is 0.035, and for star B it is 0.066. This explains the difference in settling rates shown in Fig.~\ref{fig:tracks}.
However, observations indicate that star A has a higher present-day $Z/X$ than star B. We can reconcile this apparent contradiction only by considering the uncertainties in the observational data. 

Both stars, A and B, have developed convective envelopes.
In our method, the mixing-length parameter, $\alpha$, has an evolutionary calibration meaning. In this context, it determines the stellar radius: varying $\alpha$ changes the stellar radius. This relationship arises from our Jacobian analysis, where the derivative $\partial R/\partial\alpha$ is significantly greater than $\partial L/\partial\alpha$. Although our method determines $\alpha$,  we do not assign a strong physical meaning to the resulting values.
We present different possible relationships between $\alpha_A$ and $\alpha_B$.

\begin{table*}
\centering
\caption{Calibration results: Initial compositions, mixing-length parameters, and ages for stars A and B.}
\label{tab:results_main}
\begin{tabular}{ccccccc}
\hline
Model &  $Z_{\mathrm{ini}}$ & Age (Ga) & $Y_{\mathrm{ini}}$ & $\alpha_A$ & $\alpha_B$ & Core of star A \\
\hline
1 & 0.0320  & 7.8257 & 0.2796 & 2.0285 & 2.0081 & Rad. \\
2 & 0.0350  & 7.8007 & 0.2880 & 2.0692 & 2.0254 & Rad. \\
\hline 
\end{tabular}
\tablefoot{$(M, R, $and $ L)$ from \citep{akeson2021}, $(Z/X)$ from \citep{morel2018},  high-Z solar abundances adopted from \citep{grevesse1993}.}
\end{table*}

\subsection{Sensitivity of the solution to observational uncertainties}
\label{subsec:sensitivity}

\subsubsection{Impact of luminosity, radius, and mass uncertainties}\label{subsubsec:uncertainties_LRM}

The measured and estimated values of $Z/X$, luminosity, radius, and masses have associated uncertainties. 
In the previous section, we estimated how the uncertainty in $Z/X$ affects the derived solution.
Specifically, the uncertainties of $\Delta(Z/X) = \pm0.0050$ result in an estimated stellar age uncertainty of $\Delta$Age = $\pm0.0125$ Ga.

We now assess the impact of other observational uncertainties (Table~\ref{tab:observations}) on the derived stellar parameters. 
We adopt the solution for $Z_{\mathrm{ini}} = 0.0320$ (model 1, Table~\ref{tab:results_main}) as the base model for the subsequent analysis. The results are presented in Table~\ref{tab:errors} and Appendix~\ref{sec:Appendix}. The first four rows show the results when luminosities vary by
$\Delta L_A/L_\odot = \pm0.0019$ and $\Delta L_B/L_\odot = \pm0.0007$, resulting in an estimated age uncertainty $\Delta$Age = $\pm0.0485$ Ga. 
The next four rows present the results for uncertainties in the stellar radii, $\Delta R_A/R_\odot = \pm0.0055$ and $\Delta R_B/R_\odot = \pm0.0036$. These variations produce an age uncertainty $\Delta$Age = $\pm0.1149$ Ga. Finally, the mass uncertainties, $\Delta M_A/M_\odot = \pm0.0020$ and $\Delta M_B/M_\odot = \pm0.0020$, result in significantly larger age errors of $\Delta$Age = $\pm0.6202$ Ga (see the last four rows). Thus, we conclude that our age estimate is most sensitive to uncertainties in stellar masses.

The most significant effect occurs when an input parameter is varied in opposite directions for stars A and B; for example, increasing the mass of star A while decreasing the mass of star B and vice versa. This indicates that the results are also sensitive to differences between the input parameters of stars A and B.

\begin{table*}
\centering
\caption{Influence of observational errors on obtained results.}
\label{tab:errors}
\begin{tabular}{clcccc}
\hline\hline
Model & {Disturbed target data} & {Age (Ga)} & { $Y_{\mathrm{ini}}$ } & {$\alpha_A$} & {$\alpha_B$} \\

\hline
1 & Base model                & 7.8257 & 0.2796 &  2.0285 & 2.0080 \\
3 &  $-\Delta L_A$, $-\Delta L_B$ & 7.8374 & 0.2794  & 2.0263 & 2.0058 \\
4 &  $+\Delta L_A$, $+\Delta L_B$ & 7.8139 & 0.2798 &  2.0306 & 2.0102 \\
5 &  $-\Delta L_A$, $+\Delta L_B$ & 7.7772 & 0.2801 & 2.0194 & 2.0074 \\
6 &  $+\Delta L_A$, $-\Delta L_B$ & 7.8735 & 0.2792 &  2.0377 & 2.0086 \\
7 &  $-\Delta R_A$, $-\Delta R_B$ & 7.8665 & 0.2790 & 2.0646 & 2.0608 \\
8 &  $+\Delta R_A$, $+\Delta R_B$ & 7.7792 & 0.2803 & 1.9900 & 1.9522 \\
9 &  $-\Delta R_A$, $+\Delta R_B$ & 7.7108 & 0.2807 & 2.0445 & 1.9456 \\
10 &  $+\Delta R_A$, $-\Delta R_B$ & 7.9358 & 0.2786 & 2.0104 & 2.0688 \\
11 &  $-\Delta M_A$, $-\Delta M_B$ & 7.9391 & 0.2796 & 2.0400 & 2.0108 \\
12 &  $+\Delta M_A$, $+\Delta M_B$ & 7.9127 & 0.2775 &  2.0421 & 2.0284 \\
13 &  $-\Delta M_A$, $+\Delta M_B$ & 8.4459 & 0.2741 &  2.1082 & 2.0841 \\
14 &  $+\Delta M_A$, $-\Delta M_B$ & 7.2110 & 0.2853 & 1.9500 & 1.9310 \\
\hline
\end{tabular}

\tablefoot{$\Delta L_A/L_\odot = 0.0019$, $\Delta L_B/L_\odot = 0.0007$, $\Delta R_A/R_\odot = 0.0055$, $\Delta R_B/R_\odot = 0.0036$ \citep{akeson2021}, $\Delta M_A/M_\odot = \Delta M_B/M_\odot = 0.0020$. Base is model 1 from Table~\ref{tab:results_main}. $Z_{\mathrm{ini}} = 0.0320$, $(Z/X)$ from Table~\ref{tab:observations} (high-Z) for all models in this table. All models have radiative cores.}
\end{table*}

\subsubsection{Impact of the solar mixture assumption}
\label{subsubsec:solarabund}

\cite{morel2018} report abundances in $\alpha$ Centauri relative to the solar value. However, the solar metallicity itself is subject to some uncertainty (e.g., \citealt{christensen2021}). In this section, we examine how the solar mixture assumption (low-Z versus high-Z) affects our results.

In addition to the total heavy-element mass fraction $Z$, the relative mass fractions of the individual elements differ between low-Z and high-Z mixtures. For example, oxygen is one of the most abundant heavy elements. Its logarithmic abundance is $\lg A(O) = 8.87$ in the high-Z mixture \citep{grevesse1993} and $\lg A(O) = 8.69$ in the low-Z mixture \citep{asplund2009}. The abundance of iron remains constant in both cases, $\lg A(Fe) = 7.50$, but the relative mass fractions of other individual elements differ. These differences affect the plasma opacity, modify stellar models, and impact age estimates. For example,  \citet{guillaume2024} studied this effect in the Methuselah star (HD140283).

We performed computations for $(Z/X)_A = 0.0312 \pm 0.005$ and $(Z/X)_B = 0.0300 \pm 0.005$, based on the solar ratio $(Z/X)_\odot = 0.0181$ adopted from the low-Z mixture \citep{asplund2009}. We used the corresponding opacity tables for this mixture of elements in the calculations. We find that the estimated age increases to 8.8 Ga. The initial helium abundance decreases to $Y_{\mathrm{ini}} = 0.259$--0.276. The initial mass fraction of heavy elements also decreases to $Z_{\mathrm{ini}} = 0.023$--0.027. The mixing-length parameters increase slightly, ranging from 2.2 to 2.3, but the difference between the two components remains below 0.03 (see Table~\ref{tab:results_var}, models 15 and 16).

\begin{table*}
\centering
\caption{Calibration results obtained for various input data.}
\label{tab:results_var}
\begin{tabular}{clcccccc}
\hline\hline
Model & Target data (see Table~\ref{tab:observations}) & Age (Ga) & $Z_{\mathrm{ini}}$ & $Y_{\mathrm{ini}}$ & $\alpha_A$ & $\alpha_B$ & Core of star A \\
\hline\hline
15 & $(M, R, L)^a$, $(Z/X)^b$, low-Z  & 8.7898 & 0.0230 & 0.2594 & 2.2157 & 2.2408 & Conv. \\
16 & $(M, R, L)^a$, $(Z/X)^b$, low-Z & 8.6950 & 0.0270 & 0.2756 & 2.2952 & 2.2942 & Conv. \\
\hline
17 & $(M, R, L)^c$, $(Z/X)^d$, high-Z & 9.2142 & 0.0320 & 0.2502 & 2.2514 & 2.3387 & Rad. \\
\hline
18 & $(M, R, L)^a$, $(Z/X)^b$, low-Z, ovsh=0.05 & 9.4321 & 0.0230 & 0.2554 & 2.3898 & 2.3176 & Conv. \\

19 & $(M, R, L)^a$, $(Z/X)^b$, low-Z, ovsh=0.20 & 10.9102 & 0.0230 & 0.2464 & 2.9347 & 2.5092 & Conv. \\
\hline
\end{tabular}
\tablefoot{
\tablefoottext{a}{\citet{akeson2021}},
\tablefoottext{b}{\citet{morel2018}},
\tablefoottext{c}{\citet{kervella2017}},
\tablefoottext{d}{\citet{porto2008}}.
}

\end{table*}

\subsubsection{Effect of alternative input parameters on the calibration}  
\label{subsubsec:choisemass}.

We examined the calibration results using the input parameters of \citep{kervella2017} and \citep{porto2008} in place of those of \citep{akeson2021} and \citep{morel2018}. 
 \citet{kervella2017}  report higher masses, radii, and luminosities than \citet{akeson2021}, whose values we adopted in Sect.~\ref{subsec:results}. In particular, the mass differences are $\Delta M_A = 0.0267 M_\odot$ for star A and $\Delta M_B = 0.0281 M_\odot$ for star B. These discrepancies arise from different observational methods.

As shown in Sect.~\ref{subsubsec:uncertainties_LRM}, age estimates are highly sensitive to stellar masses, particularly to the mass difference between $M_A$ and $M_B$. Other input parameters also affect the results. We examined a more complex scenario in which all input parameters vary simultaneously. We performed computations using the full set of input data -— masses, radii, and luminosities -— from \citep{kervella2017}, along with the $Z/X$ ratio from \citep{porto2008}. In this case, we assumed that the metallicities of both stars are  equal within uncertainties. For the high-Z solar mixture, we adopted the metallicity from \citep{grevesse1993}. 
We present the results in Table~\ref{tab:results_var}, model 17. The estimated age in this case is 9.2 Ga, which is 1.4 Ga older than the age derived using the input parameters from \citep{akeson2021}. Therefore, adopting the parameters of \citep{akeson2021} yields a younger age estimate for $\alpha$ Centauri.

\subsection{Acoustic properties of the stellar models} \label{subsec:acoustic}

In this section, we examine the acoustic properties of the models for star A (see Tables~\ref{tab:results_main} -- \ref{tab:results_var}). These models differ in age, in the presence or absence of a convective core, and in mass, luminosity, and radius -- parameters that were varied within the limits of observational uncertainties.

For star B, only a few frequencies have been detected, which is insufficient for a detailed study comparable to that presented here for star A. We provide a brief discussion of star B at the end of this section.

\subsubsection{Sound speed profiles in the stellar models}
\label{subsubsec:sound}

\begin{figure*}
\centering
\includegraphics[width=\columnwidth]{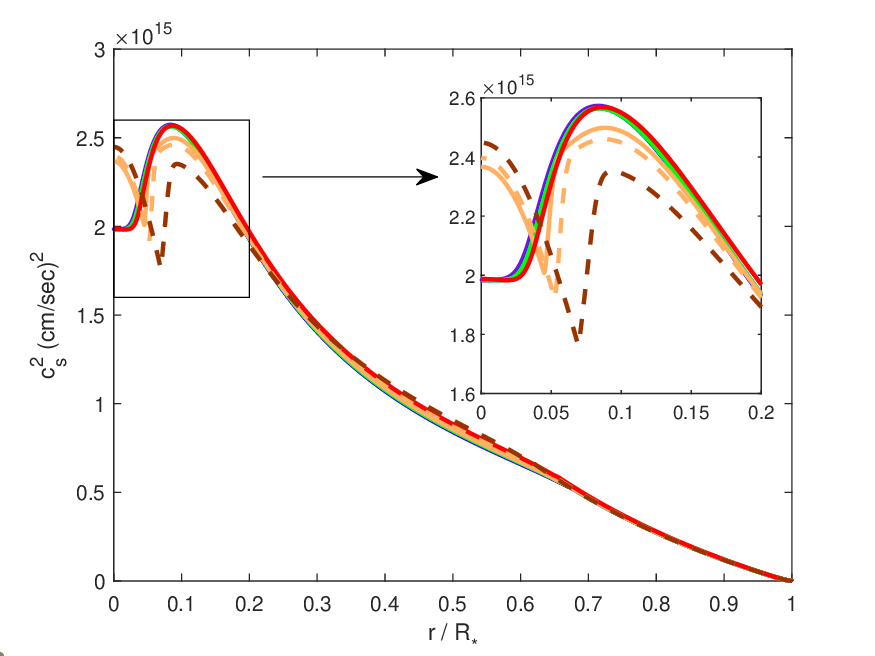}
\includegraphics[width=\columnwidth]{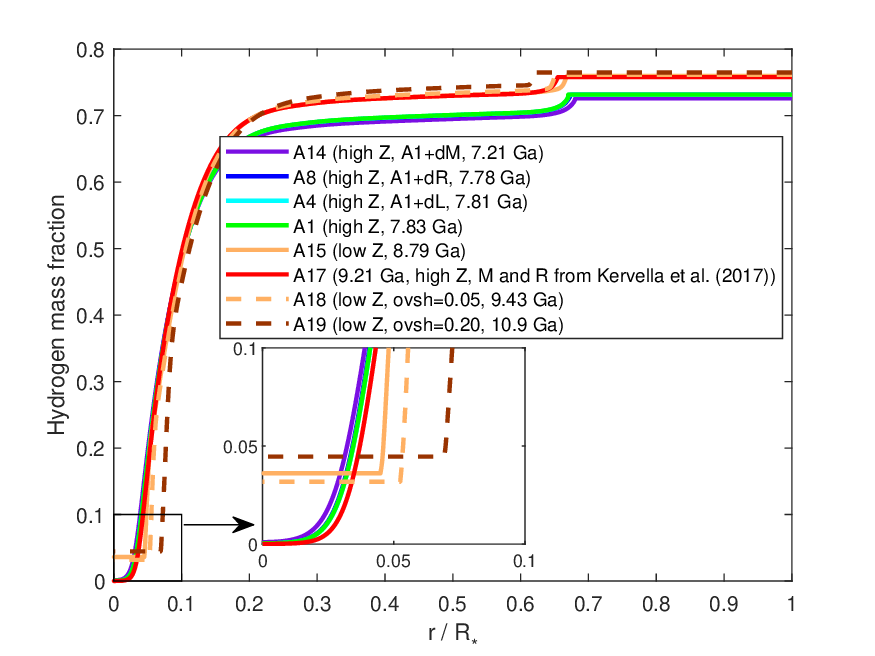}
\caption{Squared sound speed profile (left) and hydrogen mass fraction (right) for models of $\alpha$ Centauri A. Models with a radiative core are shown by solid curves, and those with convective core are shown by dashed curves.}
\label{fig:sound_hydrogen}
\end{figure*}

Figure~\ref{fig:sound_hydrogen} shows the sound speed profiles for the different models. In the convective zone, the profiles overlap, but in the radiative zone, between $r/R_*$ = 0.2 and 0.7, the sound speed in younger stars is lower than in older ones.

In the central regions, at $r/R_*<0.2$, the situation is more complex. Models divide into two clear groups: those with a radiative core and those with a convective core. 

In models with a convective core, the sound speed initially decreases with distance from the center, then rises sharply at the core boundary, and finally decreases smoothly. Models A15, A18, and A19 (Table~\ref{tab:results_var}) belong to this group and were computed using the low-Z solar mixture. Model A15 has an age of 8.8, while model A18, which includes overshooting at 0.05$H_p$, is slightly older, 9.4 Ga. Model A19 has an age of 10.9~Ga. The sound speed in the center of these models is noticeably higher than in those with a radiative core. Models that include overshooting are extensively discussed in \citet{bazot2016}.

Convective-core models retain hydrogen in their central regions, while radiative-core models have completely depleted it (Fig.~\ref{fig:sound_hydrogen}). The presence of a convective core in star A remains an open question. Supporting radiative cores, \citet{meulenaer2010} found that models with radiative cores better reproduce the observed $r_{10}$ and $d_{13}(n)$ frequency separations than those with convective cores. Similarly, \citet{bazot2016} found that 60\% of their models have radiative cores and 40\% have convective cores. \citet{nsamba2018,nsamba2019} reported that 30\% of their models have radiative cores, while 70\% contain convective cores.

\subsubsection{Asteroseismic frequencies to test the solution}
\label{subsubsec:frequencies}

Seismic frequencies were calculated on the basis of inverse-mapping stellar models using the ADIPLS \citep{christensen2008} and GYRE \citep{townsend2013} oscillation codes. The discrepancies between these two calculation methods do not exceed 0.13 µHz, which is less than 0.1\% of the absolute frequency values.

\begin{figure*}
\centering
\includegraphics[width=15cm]{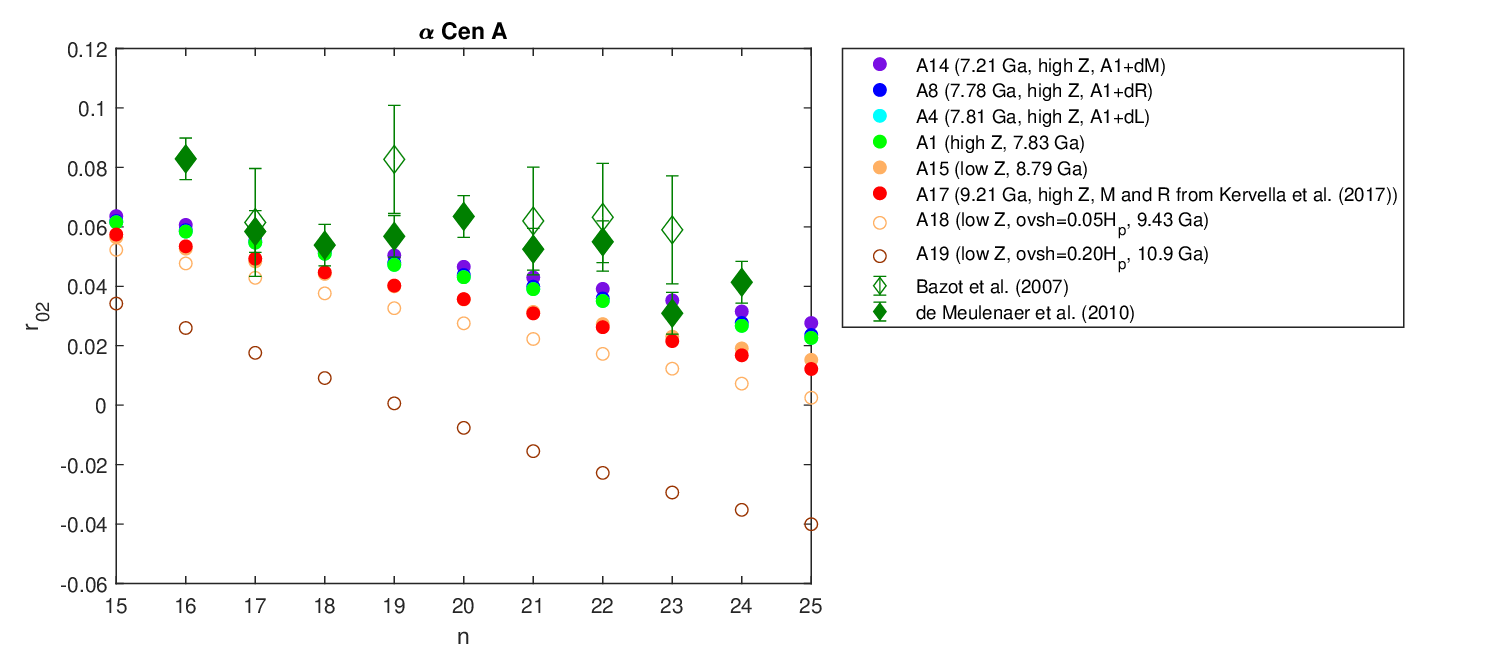}
\caption{Ratio $r_{02}(n)$ of small and large separations as function of radial order $n$ for different models of $\alpha$ Centauri A (circles). Colors indicate the stellar age and diamonds show observational data.}
\label{fig:r02}
\end{figure*}

\citet{roxburgh2003} showed that the large and small separation ratios were more convenient to analyze the stellar internal structure than the frequencies themselves, as these ratios are less sensitive to the outer layers. We consider the ratio $r_{02}$,
\begin{equation}
r_{02} = \frac{d_{02}(n)}{\Delta_1(n)},
\end{equation}

\noindent where small separation is
\begin{equation}
d_{02}(n) = \nu_{n,0} - \nu_{n-1,2},
\end{equation}

\noindent and the large separation is
\begin{equation}
\Delta_l(n) = \nu_{n,l} - \nu_{n-1,l}.
\end{equation}
\noindent Fig.~\ref{fig:r02} presents the results.

We verified whether our models are consistent with the frequencies published by \citet{bazot2007} and \citet{meulenaer2010}. We calculated $r_{02}$ from these frequencies, using the mean when two values were available for a given combination of $n$ and $l$. Figure~\ref{fig:r02} shows that our models, normalized to high-$Z$ solar abundance, with radiative cores and ages 7.2--7.8 Ga lie along the lower edge of the observed ranges. Among them, the younger model A14 (7.2 Ga) provides the closest match.

The older the star, the lower its $r_{02}$. This
trend is slightly modified when the stellar masses vary.
The mass effect depends on the mass ratio of stars A and B, which influences both the age calibration and, consequently, the frequencies. For example, the violet points represent model A14, in which we increased the mass of star A by 0.2\% while simultaneously decreasing the mass of star B by 0.2\% (see Table~\ref{tab:errors}). This change reduces the calibration age to 7.2 Ga instead of 7.8~Ga (green points, model A1), producing a slight increase in the $r_{02}$ ratio. Another example uses the stellar masses reported by \citet{kervella2017} where the age of the pair is estimated to be 9.2 Ga, leading to a significant reduction in the $r_{02}$ ratio (red points).

Adding overshooting to the A component with convective core increases the model ages in our solution. We obtain 8.79 Ga without overshooting, 9.43 Ga with overshooting of  $0.05H_p$, and 10.9 Ga with $0.20H_p$. The increasing age of the solution is the dominant effect of overshooting. Consequently, the older model at 10.9 Ga exhibits a markedly lower $r_{02}$ than the others (Fig.~\ref{fig:r02}). 

 \citet{salmon2021} studied the effect of overshooting, but on  models of completely different ages. They found that the A component with overshooting of $0.20H_p$ has an age of 6.64 Ga, which differs significantly from our model. Therefore, we cannot directly compare the models with respect to $r_{02}$.

Models A15 (orange points) and A17 (red points) have similar $r_{02}$ despite different ages. This results from the interplay of several input parameters, particularly the abundances and masses. Radius and luminosity changes within observational uncertainties minimally affect the frequencies (blue and cyan points).

\begin{figure}
    \centering
    \includegraphics[width=\columnwidth]{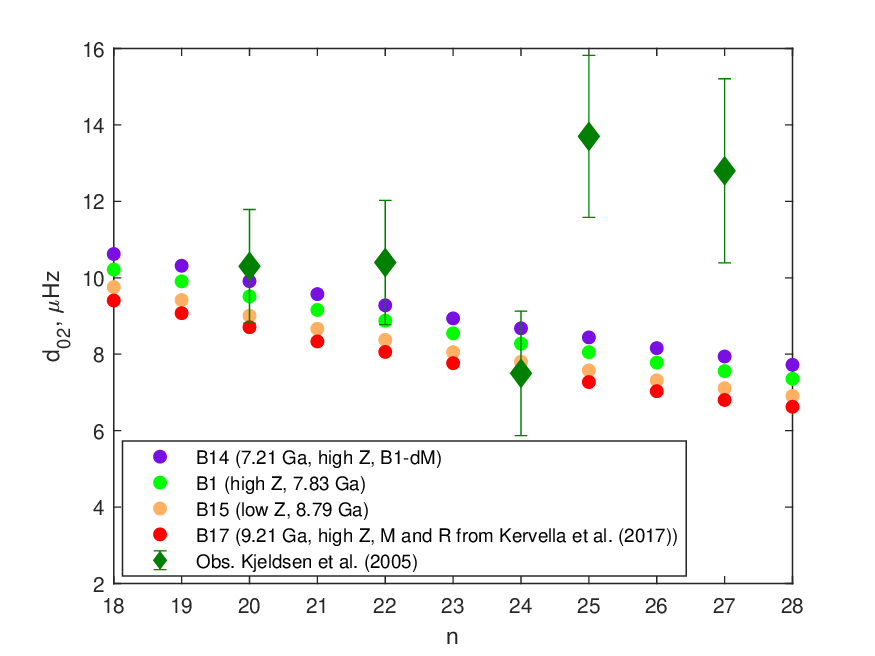}
    \caption{Small separation $d_{02}$ versus radial order $n$ for $\alpha$ Centauri B. Models are shown as circles, and observational data by \citet{kjeldsen2005} are shown as diamonds}.
    \label{fig:d02_starB}
\end{figure}

A similar discussion of the $r_{02}$ ratio for star B is not possible due to the limited detected frequencies \citep{kjeldsen2005}. Instead, we adopted a more traditional approach, comparing the observed and computed small frequency separations (Fig.~\ref{fig:d02_starB}). This analysis shows general agreement for all models of star B. 

\subsection{Comparison with previous studies}
\label{subsec:Comparison}

Stellar masses are fundamental to analyses using evolutionary tracks The adopted values for $\alpha$ Centauri A and B have evolved since the earliest studies. Over the past two decades, the estimated masses have changed by approximately 3\%, reflecting advances in observational techniques and modeling precision.
Consequently, researchers must interpret comparisons between studies with caution.

\begin{figure*}
    \centering
    \includegraphics[width=17cm]{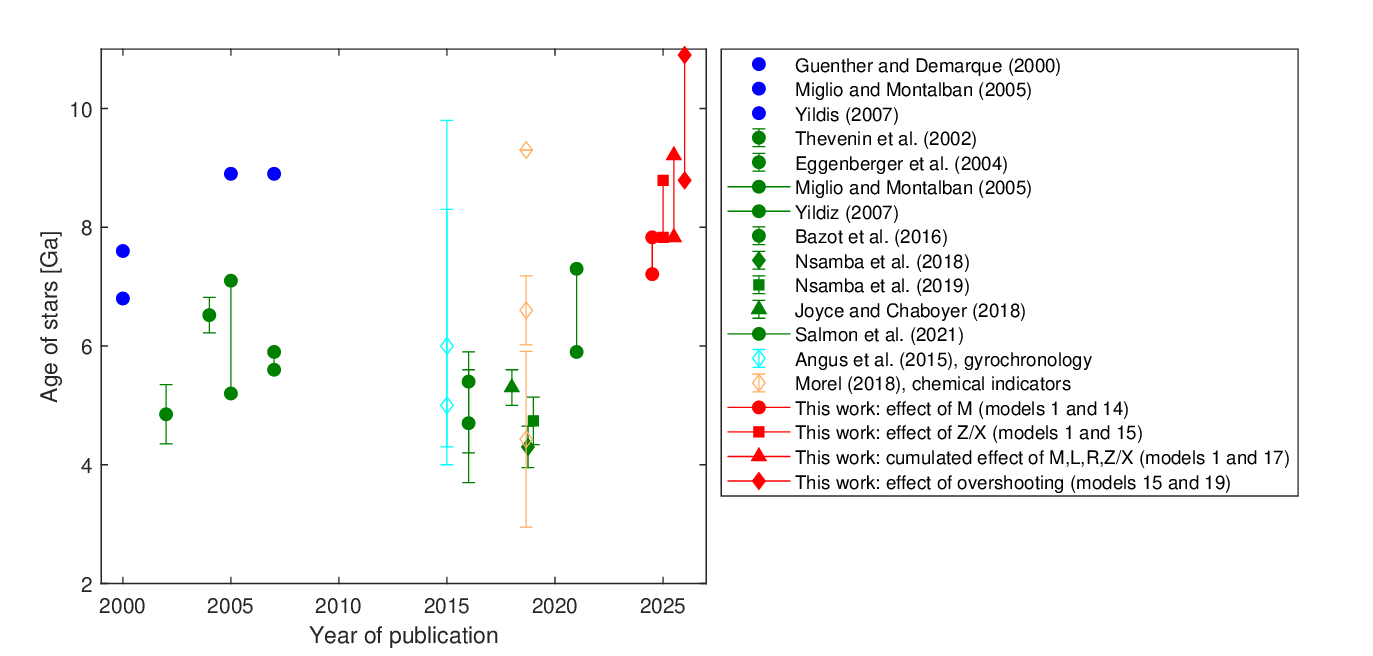}
    \caption{Age of $\alpha$ Centauri A and B estimated in various works. The colors of the points indicate the method: blue, classic method without seismic data; green,  seismic analysis; cyan, gyrochronology;  orange, chemical analysis; and red, results of this study.}
    \label{fig:other_works}
\end{figure*}

Researchers have estimated the age of $\alpha$ Centauri A and B multiple times using different methods (see Fig.~\ref{fig:other_works}). For example, \citet{angus2015} used the gyrochronology approach and estimated the age of star A ($5.0^{+3.3}_{-1.0}$ Ga) and star B ($6.0^{+3.8}_{-1.7}$ Ga) from their rotation periods. \citet{morel2018} performed a detailed abundance analysis of the spectra, deriving ages from indicators such as [Y/Mg] and [Y/Al] ( Y denotes yttrium). The ages inferred by \citet{morel2018} vary, ranging from  $6.3 \pm 1.3$ Ga for certain abundance indicators, while others yield ages exceeding 8 Ga.

The classic method to estimate the age of binary stars is evolutionary calibration without seismic data (blue points in Fig.~\ref{fig:other_works}). \citet{guenther2000} provided one of the most comprehensive studies of $\alpha$ Centauri using this approach. They estimated the age as 7.6 Ga if star A has a convective core, or 6.8 Ga if it does not.  \citet{miglio2005}, \citet{yildiz2007} estimated the age as 8.9 Ga when seismic data were not included. Our method yields 7.2--10.9 Ga (red points in Fig.~\ref{fig:other_works}).

Incorporating seismic constraints into the calibration yields younger ages (green points in Fig.~\ref{fig:other_works}). \citet{thevenin2002} first used asteroseismic frequencies to calibrate models, obtaining an age of 4.85$\pm$0.5 Ga. Similarly, when researchers accounted for seismic data, several studies reported more tightly constrained age estimates, although they used older mass values: \citet{eggenberger2004} found 6.52$\pm$0.3 Ga, \citet{miglio2005} reported a range of 5.2--7.1 Ga, and \citet{yildiz2007} estimated 5.6--5.9~Ga. Using recent masses, \citet{joyce2018} suggest an age of $5.3\pm 0.3$ Ga and  \citet{salmon2021} report an age of 5.9--7.3 Ga.

Our results differ from \citet{joyce2018} mainly due to assumptions about the different ages and initial chemical compositions of stars A and B. In Table 4 of \citet{joyce2018}, they allow differences in the initial chemical composition, with $\Delta Y_{\mathrm{ini}} \leqslant 0.02$, $\Delta Z_{\mathrm{ini}} \leqslant 0.002$, and in age, $\Delta$Age=0.02 Ga. Our approach differs: we seek solutions in which both stars share an identical chemical composition and age. However, if we allow such differences in the initial composition, our method cannot exclude 5.5--6.0 Ga as a possible solution (see Fig.~\ref{fig:inverse}).

Moreover, we excluded seismic constraints from our evolutionary calibration to maintain a general approach applicable to any stellar binary system, since asteroseismic frequency data for binaries remain relatively scarce despite their potential to significantly decrease age estimates. Instead, we used the available seismic frequencies solely to evaluate and validate our results within an asteroseismic context.

\section{Conclusions}
\label{sec:conclusion}

We developed and tested a method to map evolutionary tracks and estimate the age and initial chemical composition of stars in multiple systems, such as binaries, assuming a common evolutionary timescale. The method is based on the inversion of evolutionary tracks constrained by observational data.

We applied our method to the nearby and well-studied system $\alpha$ Centauri system to assess how uncertainties in the mass, radius, and luminosity of the two components affect the derived system age. 
Our age estimates span 7.2 and 10.9 Ga. Using the latest data \citep{akeson2021} yields an age of 7.8 Ga, which is 1.4 Ga younger than the estimate based on previous data \citep{kervella2017}. Varying the component masses within their possible errors further reduces the age to 7.2 Ga.
We investigated the chemical composition of the plasmas, focusing on the relative abundances of heavy elements for a given iron abundance. We considered two cases:  high-$Z$ and low-$Z$ solar mixtures. These assumptions produce different ages: 7.8 Ga for high-$Z$ and 8.7 Ga for low-$Z$. The high-$Z$ case produces an
isothermal, nonhydrogen core in component A. The low-$Z$ case yields an older age and a small convective core in component A. Enlarging the convective core with overshooting of $0.20H_p$ increases the age to 10.9 Ga.

Models with a convective core and significant overshooting yield ages of $9.4-10.9$ Ga. These ages appear implausible because they require a high $Z_{\mathrm{ini}} = 0.0230$, which is unlikely for old stars. By comparison, the initial abundance of heavy elements in the Sun is $Z_{\mathrm{ini}} = 0.0130-0.0188$ \citep{asplund2021}. Therefore, we favor the $7.2-7.8$ Ga age obtained from models with a radiative core.

The mass and chemical composition provide the primary data for determining the age. In addition, frequencies provide complementary information on the internal structure. They are sensitive to the central hydrogen content and thus serve as an age indicator. Improving the frequency precision for both stars can lead to more accurate age estimates. The required accuracy should be on the order of 0.2 $\mu$Hz (see Fig.~\ref{fig:d02_starB}).

To maintain a general approach, we excluded asteroseismic frequencies in the constraints for the binary system, since such data are often unavailable for binary stars. Nevertheless, asteroseismic methods can significantly enhance our  understanding of the internal structure of the two components and improve age estimates. Because asteroseismic analysis is independent of evolutionary calibration, we considered it separately in our $\alpha$ Centauri test.

Asteroseismic frequencies are available for both components and are important for refining stellar models. We used these frequencies separately to assess the influence of a convective core with overshooting on age determination and to explore their potential to further constrain the models.
Our models, computed with high-metallicity solar abundances and ages of 7.2--7.8 Ga, lie along the lower boundary of the observed $r_{02}$ parameter space. Among them, the youngest model, A14, at 7.2~Ga, provides the closest match.

Incorporating asteroseismic data into the method is a goal for future studies when such data become available. This will require reformulating the approach and could extend its application to binaries stars or stars in stellar clusters. 
The $\alpha$ Centauri system provides a rare and valuable example in which a comprehensive set of observational constraints can be applied. The method's development will be particularly significant if extensive and accurate frequency measurements become available in the future.

\begin{acknowledgements}
 We are grateful to prof. J.~Christensen-Dalsgaard for the opportunity to work with ADIPLS\footnote{\href{https://users-phys.au.dk/~jcd/adipack.v0_3b/}{https://users-phys.au.dk/$\sim$jcd/adipack.v0\underline{\hphantom{a}}3b/}}. We thank professors R.~H.~D.~Townsend and S.~A.~Teitler for the opportunity to work with the GYRE code\footnote{\href{https://gyre.readthedocs.io/en/stable}{https://gyre.readthedocs.io/en/stable}}. 
 The study by V.~A.~Baturin, A.~V.~Oreshina, S.~V.~Ayukov, and A.~B.~Gorshkov was conducted under the state assignment of Lomonosov Moscow State University.
\end{acknowledgements}

\begin{appendix}
\onecolumn
\section{Influence of disturbances on obtained results}
\label{sec:Appendix}

\begin{table*}[h!]
\centering
\caption{Influence of disturbances on obtained results. Difference between models 3-14 from Table~\ref{tab:errors} and base (model 1 from Table~\ref{tab:results_main})}
\label{tab:errors_diff}
\begin{tabular}{ccccccc|cccc}
\hline\hline
\multirow{2}{*}{Model} & \multicolumn{6}{c|}{Disturbances} & \multicolumn{4}{c}{Output} \\
\cline{2-7} \cline{8-11}
 & $\Delta L_A/L_\odot$ & $\Delta R_A/R_\odot$ & $\Delta M_A/M_\odot$ & $\Delta L_B/L_\odot$ & $\Delta R_B/R_\odot$ & $\Delta M_B/M_\odot$ & $\Delta Y_{\mathrm{ini}}$ &$\Delta$ Age (Ga) & $\Delta\alpha_A$ & $\Delta\alpha_B$ \\
\hline
3 & $-0.0019$ & 0 & 0 & $-0.0007$ & 0 & 0  & $-0.0002$ & $+0.0117$ & $-0.0022$ &  $-0.0022$\\
4 & $+0.0019$ & 0 & 0 & $+0.0007$ & 0 & 0  & $+0.0002$ & $-0.0117$ & $+0.0021$ &  $+0.0021$\\
5 & $-0.0019$ & 0 & 0 & $+0.0007$ & 0 & 0  & $+0.0004$ & $-0.0485$ & $-0.0091$ &  $-0.0006$\\
6 & $+0.0019$ & 0 & 0 & $-0.0007$ & 0 & 0  & $-0.0004$ & $+0.0478$ & $+0.0091$ &  $+0.0005$\\
7 & 0 & $-0.0055$ & 0 & 0 & $-0.0036$ & 0 & $-0.0006$ & $+0.0408$ & $+0.0360$ &  $+0.0527$\\
8 & 0 & $+0.0055$ & 0 & 0 & $+0.0036$ & 0 & $+0.0007$ & $-0.0465$ & $-0.0385$ &  $-0.0558$\\
9 & 0 & $-0.0055$ & 0 & 0 & $+0.0036$ & 0 & $+0.0011$ & $-0.1149$ & $+0.0159$ &  $-0.0624$\\
10 & 0 & $+0.0055$ & 0 & 0 & $-0.0036$ & 0 & $-0.0010$ & $+0.1101$ & $-0.0181$ &  $+0.0607$\\
11 & 0 & 0 & $-0.0020$ & 0 & 0 & $-0.0020$ & $-0.0001$ & $+0.1134$ & $+0.0115$ &  $+0.0027$\\
12 & 0 & 0 & $+0.0020$ & 0 & 0 & $+0.0020$ & $-0.0021$ & $+0.0870$ & $+0.0135$ &  $+0.0203$\\
13 & 0 & 0 & $-0.0020$ & 0 & 0 & $+0.0020$ & $-0.0055$ & $+0.6202$ & $+0.0796$ &  $+0.0760$\\
14 & 0 & 0 & $+0.0020$ & 0 & 0 & $-0.0020$ & $+0.0056$ & $-0.6147$ & $-0.0785$ &  $-0.0770$\\
\hline
\end{tabular}
\end{table*}

\end{appendix}


\begin{thebibliography}{}

\bibitem[Akeson et al.(2021)]{akeson2021} Akeson, R., Beichman, Ch., Kervella, P. 2021, AJ, 162, 14

\bibitem[Angulo et al.(1999)]{angulo1999} Angulo, C., Arnould, M., Rayet, M., and the NACRE collaboration 1999, Nuclear Physics A, 656, 3

\bibitem[Angus et al.(2015)]{angus2015} Angus, R., Aigrain, S., Foreman-Mackey, D., McQuillan, A. 2015, MNRAS, 450, 1787

\bibitem[Asplund et al.(2009)]{asplund2009} Asplund, M., Grevesse, N., Sauval, A.~J., Scott, P. 2009, ARA\&A, 47, 481

\bibitem[Asplund et al.(2021)]{asplund2021} Asplund, M., Amarsi, A.M., Grevesse, N.,  2021, A\&A, 653, A141

\bibitem[Bazot et al.(2007)]{bazot2007} Bazot, M., Bouchy, F., Kjeldsen, H., et al. 2007, A\&A, 470, 295

\bibitem[Bazot et al.(2016)]{bazot2016} Bazot, M.,  Christensen-Dalsgaard, J., Gizon, L., et al. 2016, A\&A, 460, 1254

\bibitem[Böhm-Vitense(1958)]{bohm1958} Böhm-Vitense, E. 1958, ZAp, 46, 108

\bibitem[Burgers(1969)]{burgers1969} Burgers, J.~M. 1969, Flow equations for composite gases (Academic Press, New York and London)

\bibitem[Chmielewski et al.(1995)]{chmielewski1995}  Chmielewski, Y. , Cayrel de Strobel, G. , Cayrel, R. , et al. 1995, A\&A, 299, 809

\bibitem[Christensen-Dalsgaard(2008)]{christensen2008} Christensen-Dalsgaard,
J. 2008, Ap\&SS, 316, 113

\bibitem[Christensen-Dalsgaard(2021)]{christensen2021} Christensen-Dalsgaard, J. 2021, Living Reviews in Solar Physics, 18, 2

\bibitem[de Meulenaer et al.(2010)]{meulenaer2010} de Meulenaer, P., Carrier, F., Miglio, A., et al. 2010, A\&A, 523, 54

\bibitem[Eddington(1930)]{eddington1930} Eddington, A. S. 1930, MNRAS, 90, 279

\bibitem[Eggenberger et al.(2004)]{eggenberger2004} Eggenberger, P., Charbonnel, C., Talon, S., et al. 2004, A\&A, 417, 235

\bibitem[Grevesse \& Noels(1993)]{grevesse1993} Grevesse, N., Noels, A. 1993, in Origin and Evolution of the Elements, ed. N.~Prantzos, E.~Vangioni-Flam, \& M.~Casse, 15

\bibitem[Guillaume et al.(2024)]{guillaume2024} Guillaume, C., Buldgen, G., Amarsi, A.~M., et al. 2024, A\&A, 692, L3

\bibitem[Guenther \& Demarque(2000)]{guenther2000} Guenther, D.~B., Demarque, P. 2000, ApJ, 531, 503

\bibitem[Iglesias \& Rogers(1991)]{iglesias1991} Iglesias, C.~A., Rogers, F.~J. 1991, ApJ, 371, 408

\bibitem[Joyce \& Chaboyer(2018)]{joyce2018} Joyce, M., Chaboyer, B. 2018, ApJ, 864, 99

\bibitem[Kervella et al.(2017)]{kervella2017} Kervella, P., Bigot, L., Gallenne, A., Thévenin, F. 2017, A\&A, 597, A137

\bibitem[Kjeldsen et al.(2005)]{kjeldsen2005} Kjeldsen, H., Bedding, T.R., Butler, R.P. et al. 2005, ApJ, 635, 1281

\bibitem[Manchon et al.(2025)]{manchon2025} Manchon, L., Deal, M., Marques, J.P.C., and Lebreton, Y., 2025, A\&A, 704, A79

\bibitem[Manchon et al.(2024)]{manchon2024} Manchon, L., Deal, M., Goupil, M.-J., et al. 2024, A\&A, 687, A146

\bibitem[Miglio \& Montalbán(2005)]{miglio2005} Miglio, A., Montalbán, J. 2005, A\&A, 441, 615

\bibitem[Morel(2018)]{morel2018} Morel, T. 2018, A\&A, 615, A172

\bibitem[Morel et al.(2000)]{morel2000} Morel, P., Provost, J. , Lebreton, Y. , et al. 2000 A\&A, 363, 675

\bibitem[Morel et al.(2001)]{morel2001} Morel, P., Berthomieux G., Provost, J., Thévenin, F. 2001 A\&A, 379, 245

\bibitem[Morel \& Lebreton(2008)]{morel2008} Morel, P., Lebreton, Y. 2008, Ap\&SS, 316, 61

\bibitem[Noels et al. (1991)]{noels1991} Noels, A., Grevesse, N., Magain, P., et al. 1991, A\&A, 247, 91

\bibitem[Nsamba et al.(2018)]{nsamba2018} Nsamba, B., Monteiro, M. J. P. F. G., Campante, T. L., et al. 2018, MNRAS, 479, 55 

\bibitem[Nsamba et al.(2019)]{nsamba2019} Nsamba, B., Campante, T.L., Monteiro, M.J.P.F.G., et al. 2019, FrASS, 6, 25

\bibitem[Porto de Mello et al.(2008)]{porto2008} Porto de Mello, G.~F., Lyra, W., Keller, G.~R. 2008, A\&A, 488, 653

\bibitem[Rogers \& Nayfonov(2002)]{rogers2002} Rogers, F.~J., Nayfonov, A. 2002, ApJ, 576, 1064

\bibitem[Roxburgh \& Vorontsov(2003)]{roxburgh2003} Roxburgh, I.~W., Vorontsov, S.~V. 2003, A\&A, 411, 215

\bibitem[Salmon et al.(2021)]{salmon2021} Salmon, S.~J.~A.~J., Van Grootel, V., Buldgen, G., et al. 2021, A\&A, 646, A7

\bibitem[Thévenin et al.(2002)]{thevenin2002} Thévenin, F., Provost, J., Morel, P., et al. 2002, A\&A, 392, L9

\bibitem[Townsend \& Teitler(2013)]{townsend2013} Townsend, R.~H.~D., Teitler, S.~A. 2013, MNRAS, 435, 3406

\bibitem[Yildiz(2007)]{yildiz2007} Yildiz, M. 2007, MNRAS, 374, 1264

\end{thebibliography}
\end{document}